\documentstyle[12pt]{article}
\def\by#1#2{{\displaystyle {#1}\over \displaystyle {#2}}}
\def\d{{\rm d}}
\def\<{\langle}
\def\>{\rangle}
\begin{document}

\begin{center}
{\Large \bf A Dynamical Model for Nuclear Structure Functions} \\ [1cm]
D.\ Indumathi, \\ 
{\it Institut f\"ur Physik, Universit\"at Dortmund, D 44221, Dortmund,
Germany} \\ [1cm]
Wei Zhu, \\ 
{\it CCAST (World Laboratory) P.O. Box 8730, Beijing 100080,
P.\ R.\ China} \\
and \\
{\it Dept. of Physics, East China Normal University, 
Shanghai 200062, P.\ R.\ China} \\
\end{center}

\vspace{0.2cm}

\begin{abstract}

{We construct a dynamical model for the parton distributions in a nucleus
by perturbative evolution of input distributions from a low starting
scale. These input distributions are obtained by modifications of
the corresponding free nucleon ones; the modifications being determined
by standard nuclear physics considerations. The model gives good
agreement with existing data. Its extension to the spin dependent case 
enables an estimation of nuclear modifications to asymmetries observed
in recent doubly polarised deep inelastic scattering experiments.
Although the structure functions themselves are very different from the
free nucleon ones, their ratio is insensitive to these changes. }

\end{abstract}

\vspace{0.5cm}

\section{Introduction}

The fact that the structure functions of bound and free nucleons are 
not equal is called the EMC effect \cite{EMC1}. Although this discovery
was made nearly fifteen years ago, the origin of the EMC effect is still
an open problem \cite{Arne,FS,Roberts}. Recently, accurate data on the
structure function ratios, $F_2^A(x, Q^2)/F_2^N(x, Q^2)$, has been 
obtained by various groups \cite{EMC2,data1,data2} on a host of nuclei
with mass numbers ranging from $A = 4$ (Helium) to $A = 118$ (Tin).
Combining these measurements with new data on the free proton and
deuteron structure functions, $F_2^p(x, Q^2)$ and $F_2^N(x, Q^2)$,
which are measured down to small $x$ values at {\sc hera} and {\sc nmc}
\cite{HERA,NMCdeut}, we have the nuclear structure function (per
nucleon), $F_2^A(x, Q^2)$. The
free nucleon data thus set a more stringent demand on the theory of the
EMC effect: the model should explain not only the ratios of the
structure functions but the absolute values of the nuclear structure
functions themselves. 

In this paper, we propose an extension of the dynamical model of the
proton structure function due to Gl\"uck, Reya, and Vogt (GRV)
\cite{GRV}, to the nuclear case. This model successfully fits
existing free proton data down to very small $Q^2 \simeq 0.5$--1
GeV${}^2$. It uses an input set of parton distributions at a low
starting scale, $Q^2 = \mu^2$, which are dynamically evolved using
the {\sc dglap} equations \cite{dglap} to obtain the distributions and
hence the
structure functions at larger values of $Q^2$. In our extension of this
model to the case of nuclear parton distributions, the free nucleon
input is modified due to nuclear effects; these are then evolved using
the same {\sc dglap} equations in order to obtain the corresponding
nuclear parton densities at larger $Q^2$. We also apply the model to the
case of spin
dependent density distributions. 

The main ingredients of our model (common to both the spin independent
as well as the spin dependent case) are as follows: At a low starting scale,
$Q^2 = \mu^2$, we picture the nucleon (either free or bound) as being
composed of valence quarks, gluons, and a mesonic sea. However, all
the parton
densities in a bound nucleon are different from those in a free one. This
is because of swelling of a bound nucleon, which causes a modification of
the partonic distributions. With the exception of one free parameter
(which characterises the extent of swelling), the bound-nucleon parton
distributions are exactly computable in terms of the known free-nucleon
ones. Nuclear binding effects then cause a further depletion of the
meson component of a bound nucleon. The extent of depletion is fixed
using standard inputs from nuclear physics. Both swelling and binding
have been applied before to the same problem \cite{Roberts,Jaffe,CRR};
however, we implement these effects in our model {\it in a totally
different manner}. These two effects therefore completely specify our
input bound-nucleon distributions at $Q^2 = \mu^2$. These are then
evolved as usual using the {\sc dglap} equations to give the nuclear
density distributions at any desired value of $(x, Q^2)$. A further
depletion of the small-$x$ densities occurs at the time of interaction. 
This is because parton overlap can occur with the nuclear medium,
resulting in a further modification of the structure function at the
scale determined by the interaction, viz., at $Q^2$. This is similar
in nature to the binding effect, and will be discussed in more detail
in what follows.

The prescription for modification of the parton densities in a free nucleon
in terms of nuclear effects can be straightforwardly extended to the
case of {\it spin dependent} parton densities. It is interesting to study
this as data on the spin dependent deuteron and neutron structure
functions have been obtained \cite{EMC} from deuteron and ${}^3$He
targets. The data have typically been analysed assuming that nuclear
effects are small in such measurements, and can thus be ignored. Our main
conclusion \cite{DI} is that the individual (spin independent as well as
spin dependent) structure functions undergo substantial modifications
due to nuclear effects; however, their {\it ratio}---the asymmetry---which
is the measured quantity, is largely free from these and so gives
hope that the neutron structure function may be unambiguously determined
from such a measurement. 

We find that our model explains all the
available data over almost the entire kinematic range in $(x, Q^2)$. 
We now discuss each of these effects in more detail. In Section 2, we
outline the original GRV approach to determining the free nucleon
structure functions, since this is the basis upon which our model is
constructed. In Section 3, we work out the bound-nucleon spin independent
structure functions, and demonstrate the agreement of the model
predictions with the data. In Section 4, we work out the spin dependent
case and indicate how it influences the extraction of the spin dependent
structure function; a major portion of the results contained in this
section has already appeared elsewhere \cite{DI}. Section 5 contains some
remarks on the individual parton distributions, especially the valence
distributions, while Section 6 concludes the paper. 

\section{The Parton Distributions in a Free Nucleon}

We work at a low starting scale of $Q^2 = \mu^2 \simeq 0.23$ GeV${}^2$.
At this low scale---the closest to the confinement scale at which a
perturbative description of the nucleonic parton content is still
meaningful \cite{GRV}---the nucleon is composed of valence quarks,
gluons and a mesonic sea. Defining $q_f^+$ and $q_f^-$ to be the
positive and negative helicity densities of $f$--flavour quarks in
either free or bound nucleons, we can define the following spin
independent and spin dependent combinations respectively:
$$
\begin{array}{rcl}
q_f(x) & = & q_f^+(x) + q_f^-(x) ~; \\
\tilde{q}_f(x) & = & q_f^+(x) - q_f^-(x) ~.
\end{array}
\eqno(1)
$$
A similar definition holds for the gluons as well. Given suitable
parametrisations of these densities at this scale (the so-called
``input densities''), the Altarelli Parisi ({\sc dglap}) equations
\cite{dglap} can
be used to perturbatively evolve the densities to any larger scale,
$Q^2$, of interest, where they can be compared with the data. Such a
parametrisation for the spin independent densities in the free-nucleon
case exists, is due to Gl\"uck, Reya, and Vogt \cite{GRV} (GRV), and
exhibits very good agreement with data amassed from various sources over
a very large kinematical region, and certainly in that of the EMC
effect, where $Q^2$ ranges from less than 1 GeV${}^2$ to around a 100
GeV${}^2$. We thus propose to use this parametrisation as a
starting point in our corresponding unpolarised model for bound-nucleon
parton distributions.

In principle, it is possible to use other available parametrisations
\cite{MRS,CTEQ}. However, we find the GRV most appropriate for our
purpose as each of their input densities is integrable (there are finite
number of partons at $\mu^2$). Every parton distribution is generated
dynamically by evolution from a set of valence-like inputs of the form
$$
xq_N(x, \mu^2) = N x^\alpha (1-x)^\beta P_{N, q}(x)~; \alpha > 0~,
\eqno(2)
$$ 
at $\mu^2$. For example, in the leading order ({\sc lo}) approximation with 
$\mu^2=0.23\, {\rm GeV}^2$, the valence ($u_v$, $d_v$), the total sea 
($S$), and gluon ($g$) input distributions in a free proton are given 
to be \cite{GRV},
$$
\begin{array}{rcl}
xu_v^N(x, \mu^2) & = & 1.377x^{0.549}(x)(1-x)^{3.027} P_{N, u}~, \\
xd_v^N(x, \mu^2) & = & 0.328x^{0.366}(x)(1-x)^{3.744} P_{N, d}~, \\
xS_N(x, \mu^2) & = & 2 x(\overline{u}+\overline{d})
			    = 2.40x^{0.29}(x)(1-x)^{7.88} P_{N, S}~, \\
xg_N(x, \mu^2) & = & 35.8x^{2.3}(1-x)^{4.0}~, 
\end{array}
\eqno(3a)
$$
where 
the polynomials, $P_{N, q}$, with the subscript $N$ referring to a free
proton, are, 
$$
\begin{array}{rcl}
P_{N, u} & = & 1+0.81\sqrt{x}-4.36x+19.4x^{3/2}, \\
P_{N, d} & = & 1+1.14\sqrt{x}+5.71x+16.9x^{3/2}, \\
P_{N, S} & = & 1+0.31x~.
\end{array}
\eqno(3b)
$$
The corresponding neutron densities can be obtained from isospin
symmetry, and hence the deuteron distributions as well, by averaging the
two. These densities, $q_N, g$, are valence-like in the sense that they
are integrable at this scale; this will prove an important consideration
when we extend this model to the nuclear case. At this low scale, we
have the following physical interpretation of the free-nucleon
distribution: the valence quarks and gluons are co-moving and constitute
the pure nucleonic component of the proton, while the sea quarks comprise
the mesonic component of the proton. This idea is reinforced by the
observation that the sea quark distribution is much softer than that of
the valence quarks or gluons (compare the exponent $\beta$ for the sea
densities with that for the valence or gluon densities), which could
possibly arise from a convolution model for mesons in nucleons
\cite{Sullivan}. Such a picture remains valid in the bound-nucleon case
and will be applied when describing nuclear modifications due to binding. 

The $Q^2$ evolution of these densities to leading order are given in
terms of the well-known splitting functions calculable from QCD; using
these, and the {\sc dglap} equations \cite{dglap}, the densities
$q_N(x, Q^2)$, at
any $Q^2 > \mu^2$ can be determined. These, as we have remarked before,
show good agreement with available data. 

The Gl\"uck, Reya, and Vogelsang (GRVs) \cite{GRVpol} parametrisation
contains the analogous treatment of the spin dependent densities,
$\widetilde{q}_N(x, \mu^2)$, which are used as an input in the
corresponding polarised problem. These densities are also parametrised
in a form similar to eq (2), and at the same value of $\mu^2$ (this is
necessary in order to retain the definition (1)); in fact, every spin
dependent density is a factor of the form of the {\sc rhs} of (2) times
the corresponding unpolarised density as given by the GRV set. Hence
the number of partons (of either helicity) is still finite. Specifically,
the spin dependent inputs for the {\sc lo} valence ($\widetilde{u_V}$,
$\widetilde{d_V}$), sea ($\widetilde{S}$ for the total nonstrange sea,
and $\widetilde{s}$ for the strange sea) and gluon ($\widetilde{g}$)
densities in the GRVs standard scenario \cite{GRVpol} are given by,
$$
\begin{array}{rcl}
\widetilde{u_V}(x, \mu^2) & = & 0.718 x^{0.2} u_V(x, \mu^2)~, \\
\widetilde{d_V}(x, \mu^2) & = & -0.728 x^{0.39} d_V(x, \mu^2)~, \\
\widetilde{S}(x, \mu^2) & = & -2.018 x (1-x)^{0.3} S(x, \mu^2)~, \\
\widetilde{s}(x, \mu^2) & = & 0.36 S(x, \mu^2)~, \\
\widetilde{g}(x, \mu^2) & = & 16.55 x (1-x)^{5.82} g(x, \mu^2)~.
\end{array}
\eqno(4)
$$
Our aim then is to compute nuclear effects and the consequent
modification of these free-nucleon input parton distributions in a
nucleon bound in a(n isoscalar) nucleus of mass $A$. We begin with
nuclear modifications in the case of spin independent lepton nucleus
deep inelastic scattering. 

\section{The Model: Spin Independent Case}

In order to retain compatibility with the free nucleon case, we work at
the same input scale of $Q^2 = \mu ^2 = 0.23$ GeV${}^2$. In other words,
we begin with the input distributions described by (3) and compute
modifications of every parton distribution due to nuclear swelling and
binding. We then evolve the resulting {\it nuclear} input distributions 
to the scale $Q^2$, and include the effects due to parton--nucleon
overlap to obtain the nuclear spin independent structure functions at
that scale. We now describe the swelling effect at the input scale $\mu^2$.

\subsection{The Swelling Effect}

Swelling was first discussed by Jaffe \cite{Jaffe} in the context of
6--quark bags within the bound nucleon. The presence of such structures
within the bound nucleon could substantially increase the confinement
radius. This confinement scale change (in which the input point,
$\mu^2$, is rescaled), was then re-expressed in terms of a
change of the scale variable, $Q^2 \rightarrow \xi Q^2$; these
rescaling models were then able to explain the EMC effect, especially in
the medium $x$ region \cite{CRR,JCRR}.

Here, we interpret swelling as an increase in the {\it physical} size of a
bound nucleon. This swelling only geometrically redistributes partons
inside the nucleon and does not change either the value of the dynamical
parameter, $\mu$, or the existing number of partons at $\mu^2$. Such an
interpretation and description of the swelling effect using some
universal principles was first discussed in a valon model by Zhu and
Shen \cite{Zhu}.
The relative increase in the nucleon's radius is $\delta_A$, where
$(R_N+\Delta R(A))/R_N = 1 + \delta_A$. While $\delta_A$ is not known,
its $A$-dependence can be written down:
$$
\delta_A=[1-P_s(A)]\delta_{\rm vol}+P_s(A)\delta_{\rm vol}/2~.
\eqno(5)
$$
The second term corrects for the fact that the swelling of a nucleon on
the nuclear surface is less than that of one in the interior. $P_s(A)$
is the probability of finding a nucleon on the nuclear surface; it is
known, and its value is discussed below. Here $\delta_{\rm vol}$
parametrises the swelling of the nucleon in the interior of a heavy
nucleus and is the only free parameter in our work. It is a constant for
nuclei with $A > 12$ and also for Helium (with $P_s = 1$) since they
have similar nuclear densities. We emphasise that the distortions of the
density distributions, being purely geometrical, conserve the total
parton number and momentum of each parton species (i.e., the first and
second moments of the distributions are unchanged). Furthermore, the
third moments are modified in a well-determined way.

Specifically, the first three moments of the parton distributions in a 
free ($q_N$) and bound ($q_A$) nucleon at $\mu^2$ are related by
$$
\begin{array}{rcl}
\langle q_A(\mu^2)\rangle_1  & = & \langle q_N(\mu^2)\rangle_1~, \\
\langle q_A(\mu^2)\rangle_2  & = & \langle q_N(\mu^2)\rangle_2~, \\
{\displaystyle (\langle q_N(\mu^2)\rangle_3-\langle q_N(\mu^2)
     \rangle_2^2)^{1/2} \over \displaystyle (\langle q_A(\mu^2)
     \rangle_3-\langle q_A(\mu^2)\rangle_2^2)^{1/2}} & = & 1+\delta_A~.
\end{array}
\eqno(6)
$$
The first two equations imply number and momentum conservation of 
each parton type in free and bound nuclei. The last incorporates the effect
of a size increase of the bound nucleon: according to Heisenberg's
uncertainty principle, this results in a pinching of the momentum
distribution, making the bound density distribution narrower and sharper
than the free nucleon ones \cite{Zhu}. This causes not only a depletion
of {\it all} parton densities at large- and small-$x$, but also an
enhancement at intermediate-$x$.

\begin{figure}

\vskip9truecm

\includegraphics{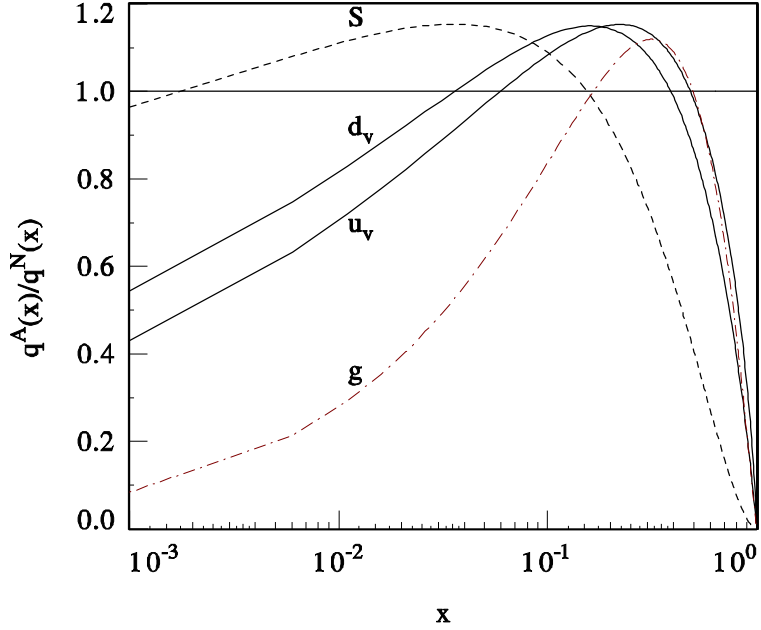}
\caption[dummy]{\small The effect of nucleon swelling on the calcium input 
distributions: the ratios of the modified to unmodified densities are
shown for $u_v$, $d_v$, $S$, and $g$. }
\label{fig1}
\end{figure}

It is clear that we can use these as constraint equations to determine
the bound-nucleon densities $q_A$ (for
$q$ = valence quarks, sea quarks, and gluons)
in terms of the free densities, $q_N$, using (3) and (6). We take $P_{A,
q}(x) = P_{N, q}(x)$, for simplicity. Then the changes in the three 
main parameters, $N$, $\alpha$, and $\beta$, in (2), due to swelling,
are immediately determined by the three constraints in (6). Thus the
input bound-nucleon densities at $Q^2 = \mu^2$ are now fixed applying
the energy--momentum constraint and the Uncertainty principle, and using
the corresponding free--nucleon densities. 
Fig.\ \ref{fig1} gives an example of the swelling effect for the 
parton densities in calcium. The momenta lost from the small and large 
$x$ regions are transferred to the intermediate $x$ region. 
The effect thus not only modifies the structure function itself, but
also weakly enhances the distributions in the region $0.1 < x < 0.3$ 
and results in ``antishadowing''. This enhancement depends on the
nuclear density (through $\delta_A$) and, furthermore, does not disappear
at larger $Q^2$ for the case of the valence densities. We will comment
on this further in Section 5. 

These modified input densities are then further modified due to the effect
of nuclear binding, which we now discuss. 

\subsection{The Binding Effect}

The bound nucleon is in an
attractive potential, $V$; $V < 0$. In traditional binding models
\cite{Roberts,LS}, the effective mass of the nucleon is taken as 
$$
M_{N,{\rm eff}}^{\rm trad} =M_N + V +T \simeq M_N+\varepsilon~,
\eqno (7)
$$
where $\varepsilon \simeq V + T $ is the average separation (or removal)
energy and the mean kinetic energy is 
$T=\langle p^2 \rangle/2M_N = 3p^2_F/10M_N$ with $p_F$ the Fermi
momentum. The average binding energy per nucleon is,
$$
b=-(\varepsilon+T)/2~.
\eqno (8)
$$
The factor 2 in (8) is due to the fact that, within the
approximation of two-body scattering, the nucleus with mass $(A-1)$
that is formed after the interaction is required to be on mass-shell.  

A consequence of (7) is a shift in the scaling variable from
$x=Q^2/(2M_N\nu)$ towards $x'=x/\langle z\rangle$ with
$\langle z\rangle =1 + \varepsilon/M_N$. This $x$-rescaling obviously
reduces the momentum of the valence quarks and explains the traditional
EMC effect (in the intermediate $x$ region). Excess mesons with energy
$E_\pi\approx T$ are therefore introduced to compensate this
energy--momentum loss. Hence the energy of the additional mesonic
fields comes
mainly from the valence quarks. Unfortunately, these kind of binding
models met with difficulty since the {\sc fnal} experiment \cite{FNAL}
excluded a net enhancement of mesonic fields in the nucleus.

In our model, we consider that the attractive potential describing the 
nuclear force arises from the exchange of mesons. At the starting
scale, $Q^2=\mu^2$, the bound nucleon is made up of valence quarks,
gluons and mesons; hence, a reasonable assumption is that the nuclear
binding effect reduces only the mesonic fields of the nucleon at this
scale. Furthermore, we regard the enhanced mesonic fields, which are a
consequence of binding, as a part of the individual nucleon itself, since
the probe cannot distinguish between the distributions of the original
mesonic fields and that of the compensative mesonic fields. The
effective mass of the bound nucleon is then
$$
M_{N,{\rm eff}} = M_N+ V + T + E_\pi = M_N - 2 b.
\eqno (9)
$$
Hence binding causes a loss of energy from the nucleons in a nucleus;
the total energy--momentum of the nucleus (which is conserved) is
obtained on adding the contribution from the total binding energy to
that  of the nucleons. 
Fermi motion further smears the parton distributions, but near
$x=1-2b/M_N$. Since we are interested here primarily in the small and 
intermediate $x$ region, we do not consider this effect in what follows.

The expression (9) allows us to simply establish the connection between the
binding effect and parton distributions in nuclei using the Weizs\"acker 
mass formula \cite{Weiz}. According to this formula, the binding energy
per nucleon arising strictly from the nuclear force is
$$
b=[1-P_s(A)]a_{\rm vol}+P_s(A)\frac{a_{\rm vol}}{2}
   = a_{\rm vol}-a_{\rm sur}A^{-1/3}~,
\eqno(10)
$$
for $A > 12$,
where $P_s(A)$ is the probability of finding a nucleon on the nuclear
surface. Experimentally $a_{\rm vol}=15.67$ MeV, and
$a_{\rm sur}=17.23$ MeV.

Coulombic interactions among nucleons will change the partonic photon
distributions and indirectly increase both the effective mass of the 
charged partons, $m_{{\rm eff}}\rightarrow m'_{{\rm eff}}$ and that of 
the nucleon, $M_{N,{\rm eff}}\rightarrow M'_{N,{\rm eff}}$. Since the 
charged partons carry most of the nucleon momentum at $Q^2=\mu^2$, 
the scaling variable $x'\simeq m'_{{\rm eff}}/M'_{N, {\rm eff}}
\simeq m_{{\rm eff}}/M_{N, {\rm eff}}\simeq x$. Therefore, we ignore 
the contributions of the Coulombic term in the mass formula.

The binding energy, $b$, corresponds to loss of energy 
of the ``mesonic'' component of the nucleons in our model. This means that  
the momentum fraction carried by the mesons in a nucleon bound in a
nucleus at $Q^2=\mu^2$ will be reduced from the free-nucleon density.
Since we have identified these to be the sea quarks in our 
model, the bound nucleon sea density is reduced from the free nucleon
one, $S_N(x, \mu^2)$, to
$$
\begin{array}{rcl}
S_A (x, \mu^2) & = & K(A)S_N(x,\mu^2) \\
 & = & \left(1-{\displaystyle 2b \over \displaystyle M_N\langle 
        S_N(\mu^2)\rangle_2}\right) S_N(x,\mu^2)~.
\end{array}
\eqno(11)
$$ 
Here $\langle S_N\rangle_2$ is the momentum fraction (second moment) of 
the mesons and the decrease in number of mesons due to the binding
effect is simply made proportional to the quark density. Since
the mesons are soft, this is a small-$x$ effect. 
The binding effect is weaker in our model since $2 b \ll - \varepsilon$;
the observed depletion of the nuclear structure function at larger $x$ 
values (the traditional EMC effect) comes from elsewhere; in fact, it
comes from modification of the densities due to swelling. 

Hence, swelling prescribes the bound-nucleon densities at the input
scale, $\mu^2$. The sea densities are additionally depleted due to
binding effects. These distributions can now be evolved to any scale,
$Q^2 > \mu^2$, using the {\sc dglap} equations. Before we use these to compute
the structure functions at this scale, however, we must incorporate one
more phenomenon. 


\subsection{The Second Binding Effect}

There is a further depletion of the sea densities which occurs at the
time of scattering, due to nucleon nucleon
interaction, arising from parton--nucleon overlap. Similar to (11), 
this causes a loss in number of sea quarks which is proportional to the
original density, i.e., 
$$
S_A(x,Q^2)-S_A'(x,Q^2) \propto S_A(x,Q^2)~.
$$
The physical origin of this depletion is easiest to see in the Breit
frame, where the exchanged virtual boson is completely space-like, so
that the 3--momentum of the struck parton is flipped in the interaction.
Hence, due to the uncertainty principle, a struck parton carrying a
fraction $x$ of the nucleon's momentum, $P_N$, during the interaction
time $\tau_{\rm int}=1/\nu$, will be off shell and localized
longitudinally to within a potentially large distance
$\Delta Z\sim 1/(2xP_N)$, which may exceed the average 2--nucleon
separation for a small enough $x < x_0$, where $x_0$ is defined below.

Binding usually occurs due to overlap of wave functions of neighbouring
nucleons. In a static nucleus, it is therefore sufficient to consider
just the binding between a nucleon and all its neighbours, and so define
an average binding energy per nucleon based on nearest neighbour
interactions. The effect of the $Q^2$ momentum transfer into the nucleus
in deep inelastic scattering is to permit overlap of wave functions
between two nucleons which
are not otherwise in contact. We parametrise this by the simple ansatz
that every overlap of the parton with a nucleon causes energy loss,
similar to binding. Hence we call this the second binding effect. The energy
loss due to this effect is from sea quarks and gluons. (Valence quarks 
are not depleted due to the requirement of quantum number conservation). 
The energy loss due to this effect from sea quarks in a bound nucleon due
to interaction with one other nucleon is therefore, 
$$
U_s(Q^2) = \beta M_N\int_{0}^{x_0}xS_A(x,Q^2)
    \simeq\beta M_N\langle S_A(Q^2)\rangle_2~,
$$
and we assume that the strength of this interaction is the same as that
due to binding, viz., 
$$
\beta=\frac{U_s(Q^2)}{M_N\langle S_A(Q^2)\rangle_2}
       =\frac{U(\mu^2)}{M_N\langle S_N(\mu^2)\rangle_2}~,
\eqno(12)
$$
$U(\mu^2)=a_{\rm vol}/6$ being the binding energy between each pair 
of nucleons. Here, we have ignored the possible $Q^2$ dependence of
$\beta$. The only r\^ole of $Q^2$ here is to provide the impulse which
allows the parton--nucleon overlap to occur. 

The shadowing begins at $x = x_0 = 1/(2M_N d_N)$. Here $d_N$ is the 
average correlation distance between two neighbouring nucleons in the 
lab frame: 
$$
d_N = P_s(A) \left[R_N+(D_S-2 R_N)\right] + 
(1-P_s(A)) \left[{\displaystyle R_N \over
 \displaystyle 2} + (D_S-2 R_N)\right]~,
\eqno(13)
$$
where $R_N$ is the nucleon radius and $D_S$ is the average
2--nucleon separation, and we have corrected for surface effects in the
usual manner. The shadowing effect saturates when the struck quark wave
function completely overlaps the nucleus in the $z$-direction, at
$x = x_A$. A parton with an intermediate momentum fraction, $x$, can
overlap $(n-1)$ other nucleons, where $n=1/(2 M_N d_N x) = x_0/x$. Due
to the applicability of the superposition principle, the loss of energy
due to interaction with each of the nucleons over which the struck quark
wave function extends, is equal and additive. The total shadowing of the
sea quarks is thus given by 
$$S'_A(x, Q^2) = K'(A) S_A(x, Q^2)~,
$$
where the depletion factor is,
$$
\begin{array}{rcll}
K'(A) & = & 1, & \hbox{when}~x > x_0; \\
& = & 1-2\beta(x_0x^{-1}-1), & \hbox{when}~x_A < x < x_0; \\
& = & 1-2\beta(x_0x_A^{-1}-1), & \hbox{when}~x < x_A,
\end{array}
\eqno(14)
$$
where $2\beta = 0.037$, $x_A=1/(4\overline{R}_AM_N)$, and 
$2 \overline{R}_A \simeq 1.4 R_A$ is the average thickness of the
nucleus. Since 
$x_0 \,{\stackrel{\displaystyle {}_<}{\displaystyle {}_\sim}} \,0.1$,
this is a small $x$ effect. We emphasize that this effect, caused by
parton--nucleon overlap, acts on the intermediate state of the
probe--target interaction and does not participate in the QCD
evolution of the initial state. Hence it cannot ``disappear'' on
evolution of the input parton densities from the input scale. However,
as $Q^2 \to \infty$, $\tau_{int} \to 0$; the overlap time (time for
interaction due to parton-nucleon overlap) goes to zero, and hence such
a nucleon--nucleon interaction cannot occur. The model is therefore best
applied to (maybe large, but) finite $Q^2$. We are now ready to write
down the bound-nucleon structure function and effect a comparison with
data. 

\subsection{The Structure Function, $F_2^A(x, Q^2)$}

The input nucleon structure function (for an isoscalar nucleus) is given by 
$$
\begin{array}{rcl}
F_2^A(x,\mu^2) & = & \langle e^2\rangle \left[\rule{0mm}{4mm} 
		    xu_v^A(x,\mu^2) + xd_v^A(x,\mu^2) 
 + K(A)xS_A(x,\mu^2) \right]~,
\end{array}
\eqno(15)
$$
where the average charge square factor is $\langle  e^2\rangle = 5/18$;
$q^A(x,\mu^2)$ incorporates the effect of swelling on every 
input parton density, $q^N(x,\mu^2)$, and $K(A)$ incorporates the
binding effect. At a larger scale $Q^2$, the second binding effect also
plays a r\^ole in modifying the sea densities. The structure function
ratio of nucleons bound in two different nuclei, $A$ and $B$, at
$Q^2>\mu^2$ is thus given by,
$$
R^{\rm AB}={\displaystyle xu_v^A(x,Q^2)+xd_v^A(x,Q^2)+ 
	 K'(A)x\hat{S}_A(x,Q^2) \over \displaystyle 
      xu_v^B(x,Q^2)+xd_v^B(x,Q^2)+K'(B) x\hat{S}_B(x,Q^2)}~,
\eqno(16)
$$
where $\hat{S}_A(x, Q^2)$ corresponds to the (leading order) evolution
\cite{dglap} of 
$K(A) S_A (x, \mu^2)$, i.e., of the swelled input sea density, depleted 
by the binding effect, to the scale $Q^2$, and $K'(A)$ incorporates the
second binding effect at the scale $Q^2$. Note that for $Q^2 > \mu^2$,
the sea density also includes a finite contribution from strange quarks.
The smearing effect of Fermi motion (at large $x$) is neglected in this
work for simplicity. Hence our results are not valid at large $x$.
Including this effect by means of the usual convolution formula
\cite{Jaffe} will give the typical depletion and then rise of $R$ as
$x \to 1$, but will not affect our small- and intermediate-$x$ results. 
	
\begin{figure}[ht]

\vskip11truecm

\includegraphics{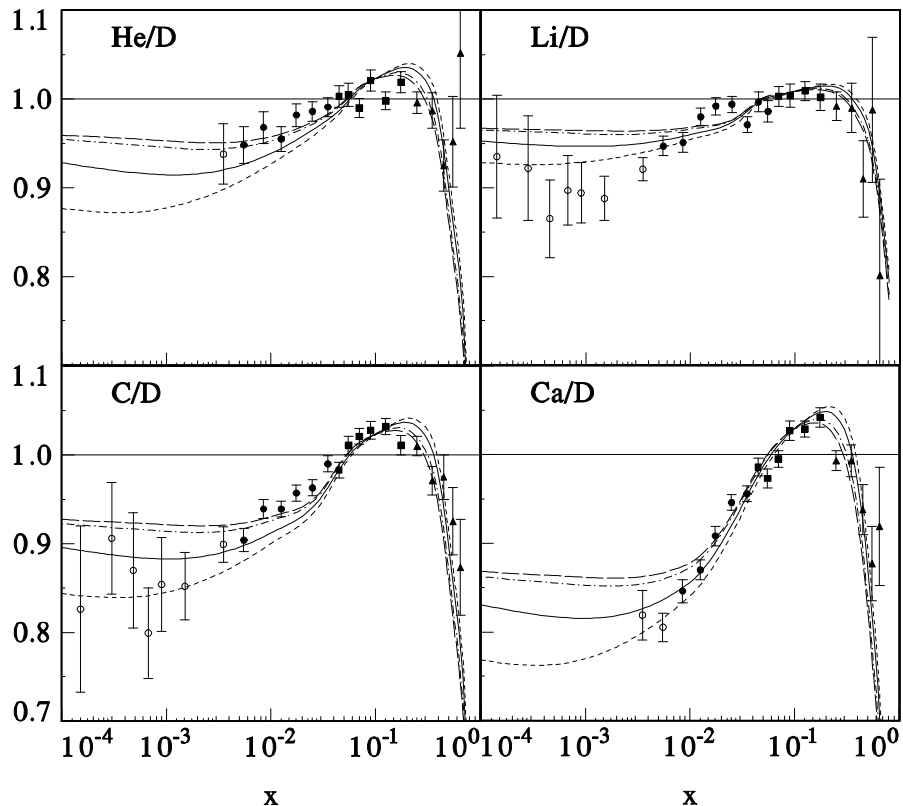}
\caption[dummy]{\small The structure function ratios as functions
of $x$ for (a) He/D, (b) Li/D, (c) C/D, and (d) Ca/D. The dashed, full,
broken, and long-dashed curves correspond to $Q^2 =$ 0.5, 1, 5, and 
15~${\rm GeV}^2$ respectively. The data \cite{data1,data2}, shown as open 
and solid circles, boxes and triangles correspond to $Q^2 <$ 1, 
1--5, 5--15, and $> 15~{\rm GeV}^2$ respectively.}
\label{fig2}
\end{figure}

In fig.\ \ref{fig2} we give our predictions for the $x$, $Q^2$, and $A$
dependences of the ratios for He/D, C/D and Ca/D. The only free parameter
to be fixed in computing these ratios is the value of
$\delta_{\rm vol}$ that parametrises the swelling. We find that
$\delta_{\rm vol} = 0.15$ best describes the various available data for
these nuclei.

\noindent We have also shown the ratio of the ${}^6$Li to D
structure functions in fig.\ \ref{fig2}. Unlike helium (which also has
$A < 12$),
lithium is a very loosely bound nucleus (whose density is less than half
that of helium or carbon, while its radius is larger than that of carbon
\cite{EMC2}). Hence, $\delta_A$ for Li can be expected to be much
smaller than that for He ($\delta_{\rm He} = 0.15/2$ according to (5))
and we take it to be 0.033. In all cases, the swelling effect dominates
the intermediate $x$ ratios while both the binding effects determine the
small $x$ ratios. 

\begin{figure}[ht]

\vskip15.5truecm
\includegraphics{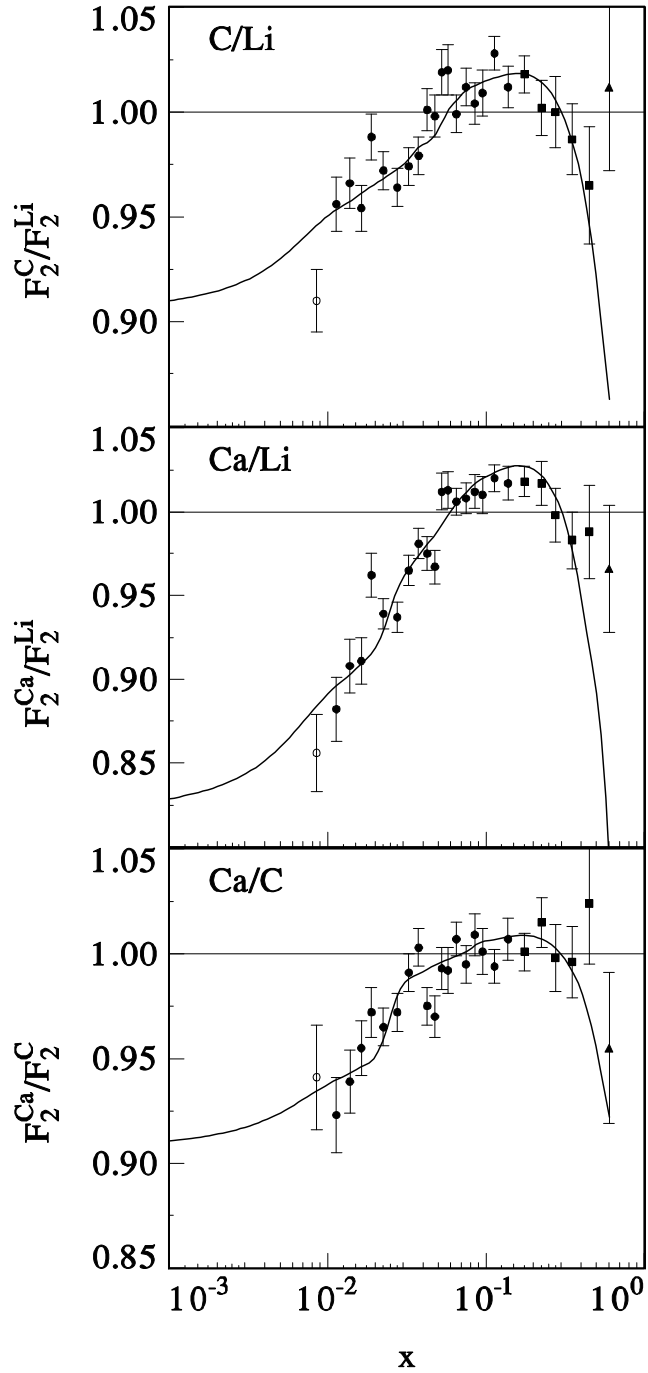}
\caption[dummy]{\small The structure function ratios for (a) C/Li, 
(b) Ca/Li, and (c) Ca/C. The solid curve is a smooth fit to our 
theoretical predictions at the same $(x, Q^2)$ as each available data 
point \cite{data1}, with $Q^2 = 0.5$ GeV${}^2$ when extrapolating to
small $x$. The description of the data is the same as in 
fig.\ \ref{fig2}.}
\label{fig3}
\end{figure}

The ratios $R^{\rm AB} = F^A_2/F^B_2$ for $A/B$ corresponding to 
C/Li, Ca/Li, and Ca/C are more sensitively dependent on the 
model of the EMC effect (as seen in fig.\ \ref{fig3}).
We predict, in general, an enhancement of the ratio,
$R^{\rm AB}$, at 
intermediate $x$ which is a net effect of the swelling and the first 
binding effect, both of which depend on the nuclear density.
On the other hand, the strong 
depletion at small $x$ depends on both the nuclear radius and density. 
The weaker enhancement of the intermediate $x$ ratios of 
Ca and C is because of their similar densities \cite{rad}
as compared to the C/Li and Ca/Li ratios where the two nuclei have 
different densities. Our predictions are seen to be in excellent 
agreement with the NMC data \cite{data1}, also shown for the
sake of comparison. 

\vspace {3cm}

\begin{figure}[ht]
\vskip9truecm
\includegraphics{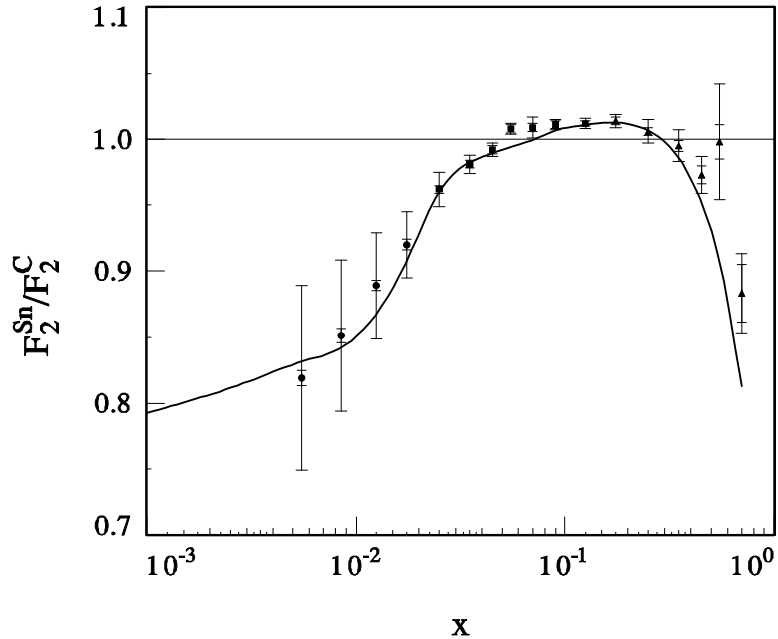}
\caption[dummy]{\small The structure function ratio for Sn/C. 
The solid curve is a smooth fit to our theoretical predictions at the
same $(x, Q^2)$ as each available data point \cite{NMCsnc}, with $Q^2 =
0.5$ GeV${}^2$ when extrapolating to small $x$. The description of the
data is the same as in fig.\ \ref{fig2}, with both statistical as well
as total (statistical and systematic, added in quadrature) errors shown.}
\label{fig4}
\end{figure}

\noindent Note that we have computed these three ratios at the
{\it same} values of $(x, Q^2)$ as the data. Finally, the NMC has
recently obtained data on the structure function ratio in Tin to Carbon
\cite{NMCsnc}. The (preliminary) data and our model predictions, again
at the same $(x, Q^2)$ as the data, are plotted in fig.\ \ref{fig4}.
We see that
there is again good agreement with the data in the region of validity of
our model ($x < 0.6$). We emphasize that, apart from the swelling
parameter, $\delta_{\rm vol}$, all parameters are fixed by a few
fundamental nuclear inputs. 

The input (3) reliably predicts the free nucleon stucture function in
the region $Q^2 \,{\stackrel{\displaystyle {}_>}{\displaystyle {}_\sim}}$
0.5 GeV${}^2$. Hence our model is also expected to predict
correctly, not only the ratios, but also the values of the nuclear structure 
functions themselves in the same kinematical region. 

We now proceeed to an analysis of the corresponding spin dependent
densities. 

\section{The Model: Spin Dependent Case}

As stated earlier, data on polarised neutron structure functions have
been obtained using deuteron and helium targets \cite{EMC}. Nuclear
effects in deuteron are known to be small (though measureable), since
the deuteron is a loosely bound nucleus. There have been a number
of papers \cite{spind} dealing with nuclear modifications of spin
asymmetries and structure functions in the case of the
deuteron\footnote{Depending on the model, corrections due to nuclear
effects in deuterium can be as large as 10\%.}. We therefore confine our
attention to possible nuclear effects on the double spin asymmetry
measurements made with helium nuclei. Woloshyn \cite{Wolo} showed that
here the protonic contribution to the asymmetry
is negligible so that the ${}^3$He double spin asymmetry is sensitive
to the spin dependent neutron structure function, $g_1^n(x, Q^2)$, alone.
However, there may be additional modifications of the spin dependent
neutron structure function itself due to nuclear effects, as has been
studied in various papers \cite{DI,CSPS}. These are
especially of importance for checking the validity of the Bjorken Sum rule
\cite{FGS} that relates the difference of the first moments of the $p$
and $n$ spin dependent structure functions to the axial vector constant
in nucleon beta decay. We therefore study nuclear effects on the spin
dependent structure functions in our model. 

The quantity of interest that is measured in polarised deep inelastic
scattering measurements is the double spin asymmetry, 
$$
{\cal A}^A (x, Q^2) = \by{g_1^A(x, Q^2)}{F_1^A(x, Q^2)}~,
\eqno(17)
$$
where $g_1$ and $F_1$ are the spin dependent and spin independent
structure functions corresponding to the nucleus $A$. (We ignore the
other structure function, $g_2$, here). We are therefore interested
in studying possible deviations, due to nuclear effects, of the measured
asymmetry, ${\cal A}^{\rm He}$, from the neutron asymmetry, ${\cal A}^n$,
that is required to be extracted from the helium data.

Recall that the study of unpolarised bound nucleon densities used the
GRV \cite{GRV} density parametrisation as an input. We therefore use
the corresponding GRVs \cite{GRVpol} spin dependent densities
as an input in the corresponding polarised problem. We proceed exactly
as in the spin independent case. 

\subsection{The Input Spin Dependent Nuclear Distributions}

The same nuclear effects of binding and swelling affect the spin
dependent densities also. This is because they influence the positive
and negative helicity densities, out of which the spin independent and
spin dependent densities are composed (see eq (1)). The entire swelling
effect can now be rephrased as the effect of swelling on individual
{\it helicity} densities, so that equations analogous to (6) are valid
for the spin dependent densities, $\tilde{q}(x)$, as well. This can be
seen as follows: Swelling simply rearranges the parton distributions in
the bound nucleon; there is no change in the number of each parton
species. In particular, each helicity type is also conserved, i.e., 
$$
\int q_A^+ (x, \mu^2) \d x= \int q_N^+ (x, \mu^2)\d x ~, ~~~~~~~~
\int q_A^- (x, \mu^2) \d x= \int q_N^- (x, \mu^2)\d x ~. 
$$
Hence, their sum and difference is also conserved. The former (the spin
independent combination) is contained in the first equation of the
equation set (6); the latter implies, for the polarised combination,
$$
\langle \widetilde{q_A} (\mu^2) \rangle_1 = 
\langle \widetilde{q_N} (\mu^2) \rangle_1~.
\eqno(18a)
$$
(Note that $\langle q(\mu^2)\rangle_n = \langle q^+(\mu^2)\rangle_n
+ \langle q^-(\mu^2)\rangle_n$ for every moment, $n$, for both the free
and bound nucleon, and similarly for the spin dependent combination as
well). Similarly, since the momentum carried by each helicity density is
unchanged, momentum conservation between the free and bound nucleon also
holds for the sum and difference of the helicity densities. The
corresponding equation for the sum is the second equation in (6); the
equation for the helicity difference is
$$
\langle \widetilde{q_A} (\mu^2) \rangle_2 = 
\langle \widetilde{q_N} (\mu^2) \rangle_2~.
\eqno(18b)
$$
The extension of the third of the equations in (6) to the spin dependent
case is not as straightforward. Every helicity density, $q_f^h(x)$, 
$(h = +, -)$, spreads out over a larger size, or, equivalently, gets
pinched in momentum space, according to Heisenberg's uncertainty
relation, $\Delta p \Delta x = 1$. Applying this to each helicity type,
for each flavour, we have, 
$$
{\displaystyle (\langle q_N^+(\mu^2)\rangle_3-\langle q_N^+(\mu^2)
     \rangle_2^2)^{1/2} \over \displaystyle (\langle q_A^+(\mu^2)
     \rangle_3-\langle q_A^+(\mu^2)\rangle_2^2)^{1/2}} = 1+\delta_A~;
     ~~~~~
{\displaystyle (\langle q_N^-(\mu^2)\rangle_3-\langle q_N^-(\mu^2)
     \rangle_2^2)^{1/2} \over \displaystyle (\langle q_A^-(\mu^2)
     \rangle_3-\langle q_A^-(\mu^2)\rangle_2^2)^{1/2}} = 1+\delta_A~.
\eqno(19)
$$
However, for later convenience, we prefer to use analogous expressions
for the sum and difference, $q_f$ and $\tilde q_f$, rather than for the
individual helicity densities. Hence, the third of the constraints
arising from swelling, i.e., the third of eq (6) and its spin dependent
counterpart read,
$$
{\displaystyle (\langle q_N(\mu^2)\rangle_3-\langle q_N(\mu^2)
     \rangle_2^2)^{1/2} \over \displaystyle (\langle q_A(\mu^2)
     \rangle_3-\langle q_A(\mu^2)\rangle_2^2)^{1/2}} = 1+\delta_A~;
     ~~~~~
{\displaystyle (\langle \widetilde{q_N}(\mu^2)\rangle_3-\langle 
\widetilde{q_N}(\mu^2) \rangle_2^2)^{1/2} \over \displaystyle (\langle 
\widetilde{q_A}(\mu^2) \rangle_3-\langle \widetilde{q_A}(\mu^2)
\rangle_2^2)^{1/2}} = 1+\delta_A~.
\eqno(18c)
$$
The error involved between the exact expressions, eq (19), and their
approximations, eq (18c), is a term proportional to $(1 - (1 + \delta_A)^2)$
and is of order $\delta_A$. In principle, it is possible to retain this
term and compute the effects of swelling in both the spin independent
and spin dependent sectors. However, this term mixes spin dependent
and spin independent moments and thus implies that spin dependent
moments affect spin independent bound-nucleon densities and vice versa,
which is certainly unappealing; alternatively, since $\delta_A$ is small
(about 10\%), these error terms are small, and can be ignored. We prefer
to take this latter approach. We are therefore justified in
using eq (18c) rather than eq (19) to constrain the second moments of
the parton densities. The three sets of equations, (18a--c), thus provide
the three sets of constraint equations, analogous to the set (6), with
which we can fix the input bound-nucleon spin dependent densities. 

The modified input densities are thus determined, given a set of valid
input free nucleon distributions, which we take to be the Gl\"uck,
Reya, and Vogelsang 'standard' set (GRVs) \cite{GRVpol} (see eq (4)).
Once again, there are three constraint equations
which serve to fix the three main parameters, $\alpha$, $\beta$, and $N$
for the corresponding bound-nucleon spin dependent densities. Note that
$\delta_A$ is the same for the unpolarised as well as the polarised case.
The effect of swelling on the spin dependent densities is thus similar
to that on the spin independent ones, causing a narrowing of the
density distributions, with resulting anti-shadowing in the
intermediate-$x$ region.

Binding causes loss of energy in the sea: this is due to loss of mesons
from the nucleon. Since these mesons are spin-0 bosons, it is clear
that no spin is lost from the sea due to binding (equal numbers of
positive and negative helicity partners are lost). Hence we see that
binding changes the sum, but not the difference of the helicity
densities\footnote{It is possible that $\rho$, etc., mesons also
participate in this interaction, leading to a change in the polarised
sea densities, but this component is small and we neglect it.}. Hence,
the only modification of the spin dependent densities at the input scale
occurs due to swelling. These are then evolved to the scale of interest.

At the time of interaction, the second binding effect applies to struck
partons with momentum fraction $x \le x_0$, as in the unpolarised case. 
The mechanism for this depletion is independent of the helicity of the
quark, and so this effect is identical in both the spin independent as
well as spin dependent cases.

\subsection{The Structure Function, $g_1^A(x, Q^2)$}

The spin dependent structure function, $g_1^A(x, Q^2)$ can now be
computed by evolving the modified input spin dependent densities to the
required value of $Q^2$, and including the second binding effect. 
We display the equivalent neutron bound-nucleon structure
functions at an arbitrary scale, $Q^2 > \mu^2$ (with $R =
\sigma_L/\sigma_T = 0$) :
$$
\begin{array}{rcl}
$$
F_1^{n/A}(x,Q^2) & = & \by{1}{18} \left[u_v^A(x,Q^2) + 4\, d_v^A(x,Q^2)
	               + K'(A) S_A(x,Q^2) \right]~, \\
g_1^{n/A}(x,Q^2) & = & \by{1}{18}  \left[\widetilde{u_v}^A(x,Q^2) + 
		        4\, \widetilde{d_v}^A(x,Q^2) 
                        + K'(A) \widetilde{S}_A(x,Q^2) \right]~.
\end{array}
\eqno(20)
$$
$q^A(\widetilde{q}^A)(x,\mu^2)$ incorporates the effect of swelling on
every input parton density, $q^N(\widetilde{q}^N)(x,\mu^2)$, as well as
that of binding for the unpolarised densities, and the corresponding
$Q^2$-dependent quantities that appear here are these input densities,
evolved suitably to the required scale.
$K'(A)$ incorporates the second binding effect at $Q^2$, as discussed
above. The experimentally measured asymmetry, and quantity of interest,
are the ratios, at the scale $Q^2$, for the neutron bound in the helium
nucleus and for a free neutron:
$$
{\cal{A}}^{\rm meas} = \by{g_1^{\rm n/He}}{F_1^{\rm n/He}}~, \hspace{3cm} 
{\cal{A}}^{\rm free} = \by{g_1^{\rm n}}{F_1^{\rm n}}~,
\eqno(21)
$$
and can thus be computed. We use the free and bound nucleon {\it
unpolarised} structure functions as computed in the previous section.
Note that the input spin dependent densities (which are taken from
\cite{GRVpol}) were actually fitted to both the free proton as well
as deuteron and ${}^3$He spin dependent data; however, we use them here as
the free nucleon parametrisations (which is permissible especially in
view of the large error bars on presently available data). Furthermore,
the smearing effect of Fermi motion (at large $x$) is neglected as in
the unpolarised case. Hence our results are not valid at large $x$. 
	
\begin{figure}[ht]

\vskip9truecm

\includegraphics{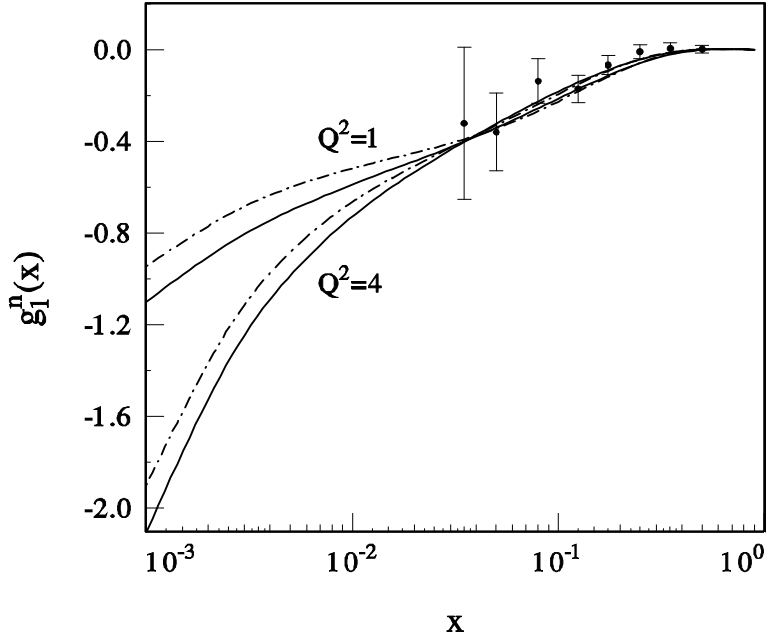}
\caption[dummy]{\small The free and bound nucleon spin dependent structure
function for $Q^2 = 1, 4 $ GeV${}^2$ as a function of $x$ are shown as
solid and dashed lines respectively. The structure function data are
extracted at $Q^2 = 4$ GeV\/${}^2$ from the asymmetries measured by the 
E142 collaboration. }
\label{fig5}
\end{figure}

In fig.\ \ref{fig5} we give the results of our computations for the
measured
(bound nucleon) and required (free nucleon) spin dependent structure
function, $g_1^n$ for typical values of $Q^2$, $Q^2 = 1, 4$ GeV${}^2$.
We see that the deviations of the bound neutron structure function can
be as large as 10--15\% at small $x$ and about 6\% at intermediate $x$
values. The data points plotted on this graph correspond to the values
extracted at $Q^2 = 4$ GeV${}^2$ from a measurement of the asymmetry by
the E142 Collaboration \cite{spind} (with $R = 0$) and indicate the size
of the error bars in currently available data.

\begin{figure}
\vskip9truecm
\includegraphics{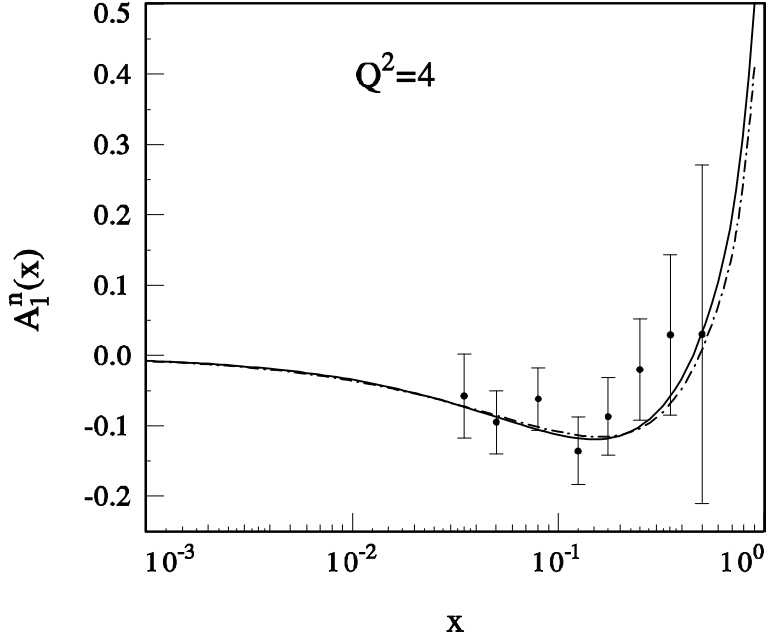}
\caption[dummy]{\small The bound and free nucleon asymmetries for $Q^2 = 4 $
GeV${}^2$ as a function of $x$ are shown as solid and dashed lines
respectively. The data are from the E142 collaboration.}
\label{fig6}
\end{figure}

In fig.\ \ref{fig6}, we plot the asymmetries at
$Q^2 = 4$ GeV${}^2$. The
data points here correspond exactly to the E142 data and therefore go over
a range of $Q^2$ with a mean of about 2 GeV${}^2$; however, the
asymmetry is not very sensitive to $Q^2$ in the $x$ range of the
available data. Notice that in this case, nuclear effects cause not more
than 5\% deviation in the asymmetry at both small and intermediate
values of $x$. The deviation is slightly larger at larger $x$, $x > 0.4$,
but this is due to the fact that the neutron spin dependent structure
function changes sign near this value, and hence this deviation cannot
be considered to be significant. 

The asymmetry can also be expressed in terms of the EMC ratios,
$$
R^A = \by{F_2^{n/A}}{F_2^n}~; ~~~~~~~~~
\widetilde{R}^A = \by{g_1^{n/A}}{g_1^n}~,
\eqno(22)
$$
for the spin independent and spin dependent structure function ratios
respectively. We have, for the helium case, 
$$
{\cal{A}}^{\rm meas} = \by{g_1^{\rm n/He}}{F_1^{\rm n/He}}
 = \by{\widetilde{R}^A}{R^A} \cdot {\cal{A}}^{\rm free}~.
\eqno(23)
$$
Our results therefore indicate that 
$${\cal{A}}^{\rm meas} \simeq {\cal{A}}^{\rm free}~,
$$
or, equivalently, that 
$$\widetilde{R}^A \simeq R^A~,
\eqno(24)
$$
over most of the observed $x$ region. Another (more experimentally
relevant) way of stating this is to examine the difference between the
following spin dependent structure functions:
$$
\begin{array}{rcl}
g_1^n & = & {\cal{A}}^{\rm free} F_1^n~; \\
g'_1 & = & {\cal{A}}^{\rm meas} F_1^n~. 
\end{array}
\eqno(25)
$$
The latter is actually what is extracted from polarised deep inelastic
scattering data while the former is the theoretical quantity of
interest. To a good approximation, however, these are numerically the
same, due to (23) and (24). Finally, it may be remarked that, although
$F_2^{n/A}$ is not known, the unpolarised bound-nucleon {\it isoscalar}
structure functions are known; hence it {\it is} possible to extract the
potentially interesting quantity, $g_1^A(x, Q^2)$, from future polarised
experiments on isoscalar nuclear targets ({\sc hermes}, for instance,
is a possibility) using,
$$
g_1^A = {\cal {A}}^{\rm meas} \cdot F_1^A~.
$$
It would then be possible to study the nuclear dependence of {\it
polarised} structure functions, which would certainly shed a great deal
of light on the nature of the EMC effect. 

In short, we see that nuclear effects in the polarised
helium data, though significant, equally
affect both the spin dependent as well as the spin independent structure
functions in such a way that the measured asymmetries are to a great
extent independent of them. Since it is the asymmetry rather than the
structure function which is measured in a polarised experiment, much
smaller errors on data are required before these small deviations due to
nuclear effects become observable in such experiments. On the other
hand, as already stated, this seems to make possible clean and
unambiguous extraction of the relevant {\it free nucleon} structure
functions from a measurement of double spin asymmetries with such
light nuclear targets.

\section{The Valence Densities}

So far we have made a comparison of the structure functions with the
data, and found good agreement. 
Our model parametrises the effects of nuclear interactions in a simple
manner to obtain the bound nucleon densities in a nucleus; however,
there are as many as three different effects that determine the
densities, and the corresponding structure functions. The valence
densities constitute the cleanest sector, as they are influenced by
exactly one nuclear effect, viz., swelling.
They thus form a good test of the {\it detailed} corectness of our
model in terms of its individual component parts. The valence sector is
already tested by current data in the large $x$ region, where the
structure function is dominated by the valence densities. We now look at
the valence behaviour in the entire $x$ region. It is possible to study
this in either charged current deep inelastic processes (by studying the
nuclear behaviour of the non-singlet structure function, $F_3(x, Q^2)$)
\cite{Kumano}, or, what is perhaps experimentally more feasible, in semi
inclusive hadroproduction in the neutral current process \cite{DIA,FSL}.
The quantity of interest here is the ratio of the difference in rates of
production of the positive and negative charged hadron in a nucleus and
a free nucleon respectively. It is defined to be
$$
R^{{\rm val}, A} = \by{\left[{{\cal N}^{h^+} - {\cal N}^{h^-}}
		    \right]^A}{\left[{\cal N}^{h^+} -
		    {\cal N}^{h^-}\right]^N}~,
$$
where ${\cal N}^{h^+}$ (${\cal N}^{h^-}$) is the production rate
(arbitrarily differential with respect to various kinematical variables)
of a hadron (its charge conjugate). Here $N$ refers to an
``isoscalar nucleon'' and $h$
refers to pions, kaons, or protons, for example. The advantage of
defining such a ratio is that the unknown fragmentation functions cancel
out in the ratio, within the approximation that the fragmentation of the
final state hadron is independent of the initial deep inelastic
scattering process (factorisation) and depends only on the momentum
fraction of the parton which fragments into the hadron. 

\begin{figure}[ht]
\vskip8truecm
\includegraphics{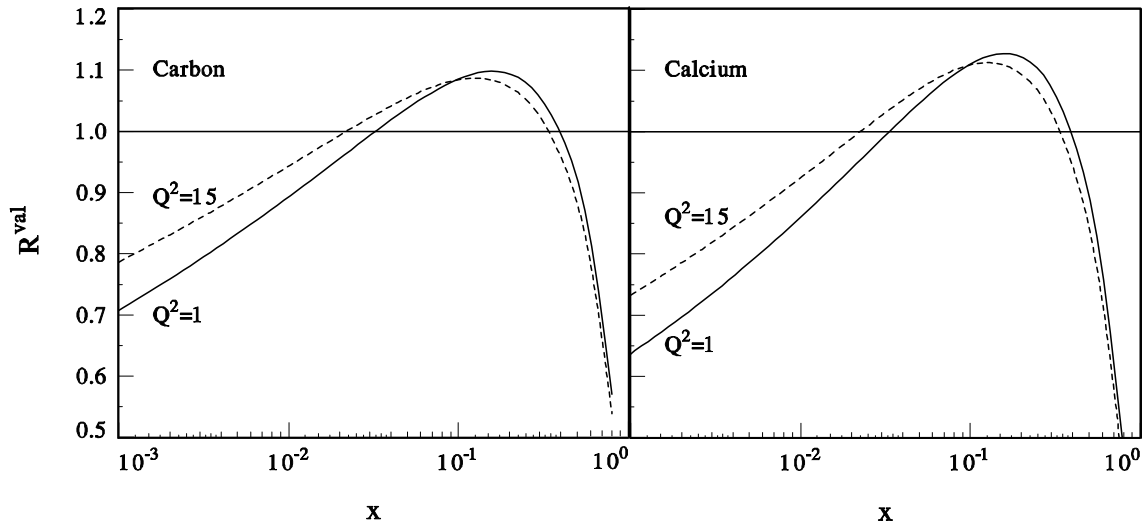}
\caption[dummy]{\small The valence ratios, $R^{{\rm val}, A}$, for
Carbon and Calcium at two different values of $Q^2$, plotted as a
function of $x$.}
\label{fig7}
\end{figure}

More specifically, the ratio in pion or kaon production from an
isoscalar nucleus, $A$, is
$$
R^{{\rm val}, A} = \by{\left(u_V^A + d_V^A\right)}
		      {\left(u_V^N + d_V^N\right)}~.
\eqno(26)
$$
Hence such a measurement yields information on just the valence sector
(the isoscalar combination) alone, and over the entire $x$ region. We
therefore plot this ratio for two different nuclei, Carbon, and Calcium,
as a function of $(x, Q^2)$ in fig.\ \ref{fig7}. The large $x$ ratio
(which is
essentially the same as the ratio of the corresponding structure
functions) is already in agreement with data; the smaller $x$ ratio
exhibits shadowing, which decreases with increasing $Q^2$, but is still
substantial at $Q^2 = 15$ GeV${}^2$.

The NMC data at low $x$ correspond
to rather small values of $Q^2$; hence, a study of semi-inclusive 
hadroproduction in such fixed target experiments should show large
shadowing in the valence sector. This is in direct contrast to rescaling
models which exhibit depletion at medium $x$ (in the traditional EMC
effect region) and antishadowing at smaller $x$ values in the valence
sector \cite{Kumano}. This is because, as we have discussed earlier,
binding causes a depletion of the {\it valence} densities in such models
at larger $x$, forcing a corresponding antishadowing at smaller $x$
(in accordance with momentum conservation). In contrast,
binding depletes the {\it sea} densities in our model, while the effect
of swelling causes depletion of the densities at both small and large
$x$ values. We remark that the aligned jet model also predicts a
shadowing \cite{FSL} of the valence ratio in (26), which is similar to
the one we obtain; however, here the depletion is due to multiple
scatterings. Hence, a measurement of the valence bound-nucleon
densities in the small $x$ region will certainly discriminate between
some of the models that are currently available. 

It is also possible to make a similar measurement in the {\it spin
dependent} sector; it turns out that the corresponding ratio, for
instance, for helium again, is not very different from the unpolarised
result. This is because the valence asymmetry, which is the
analogue of (23) for the valence quarks alone, is again largely
insensitive to $A$, just as in the case of the asymmetry with respect to
the total structure functions.
Hence, shadowing at small $x$ of the valence structure functions is
predicted for both the polarised and unpolarised measurements. 

\section{Conclusion}

We have constructed a model to compute nuclear modifications
of nucleon structure functions and have used it to compute spin
independent as well as spin dependent bound nucleon structure functions.
The model uses both swelling and binding effects and is in good
agreement with existing data. The swelling effect causes a modification
of the bound-nucleon structure function itself, rather than causing just a 
scale change, as in traditional models of the nuclear EMC effect, and
effectively gives rise to the observed intermediate-$x$ ``antishadowing''.
The modified parton distributions are completely specified in terms of the
free distributions, using fairly general conditions such as
energy--momentum constraints and the Uncertainty principle, and using
just one free parameter for the entire set of nuclei with
different mass-numbers, $A$. Binding effects cause a depletion of
the sea (or mesonic) densities in a bound nucleon. The extent of
depletion is completely determined in terms of fundamental nuclear
parameters. Both swelling and binding thus determine the input
bound--nucleon densities, which are then dynamically evolved to the scale
of interest. Finally, a second binding effect, caused by parton--nucleon
overlap at the time of interaction, is introduced. This causes a further
depletion of the small-$x$ densities; the binding effects together
account for the observed ``shadowing'' of the small-$x$ structure 
functions. Hence our model satisfactorily explains the observed EMC 
effect with remarkably few input parameters. In the case of polarised
scattering, the model predicts that the measured asymmetries may not be
sensitive to the nuclear modifications. Finally, a measurement of
semi-inclusive hadroproduction in deep inelastic scattering experiments
with nuclear targets, or the more ambitious one with polarised nuclear
targets, will give information on nuclear modifications of valence
densities in a different kinematical region as that probed by the
corresponding inclusive process.

In short, we have a comprehensive model of nuclear structure functions
that uses well-known parametrisations of the free nucleon parton
densities and modifies them due to nuclear effects in a deterministic
manner. The nuclear input is kept simple, with few free parameters, and
the agreement of the model with data is satisfactory. Just as the
general properties of the nuclear force originated from the binding
effect of nuclei in the history of nuclear physics, we expect that a
new understanding of binding effects in the EMC effect will bring to
light the nature of the nuclear force at the level of quarks and gluons. 

{\bf Acknowledgements}:
We thank M.\ Gl\"uck and E.\ Reya for extensive discussions and
continuous encouragement, as well as for a critical reading of the
manuscript. We thank M.~Stratmann for providing the relevant {\sc fortran}
programs. One of us (W.\ Z) acknowledges the support of the
DAAD--K.\ C.\ Wong Fellowships and the National Natural Science Foundation
of China and the hospitality of the University of Dortmund where part of
this work was done.


\begin{thebibliography}{99}
\bibitem{EMC1} J. J. Aubert {\it et al.}, The EMC, Phys. Lett. B123
(1982) 275. 

\bibitem{Arne} For a recent review, see 
M. Arneodo, Phys. Rep. 240 (1994) 301.
Also see references in the earlier reviews of \cite{FS}
and \cite{Roberts} below. 

\bibitem{FS} L.L. Frankfurt and M.I. Strikman, Phys. Rep. 160
(1988) 235. 

\bibitem{Roberts} R.G. Roberts, {\it The structure of the proton},
Cambridge University Press, Cambridge, 1990, Chapter 8. 

\bibitem{EMC2} P. Amaudruz {\it et al.}, The EMC, Z. Phys. C53
(1992) 73.

\bibitem{data1} P. Amaudruz {\it et al.}, The NMC, Nucl. Phys. B441 (1995)
3.

\bibitem{data2} M. Arneodo {\it et al.}, The NMC, Nucl. Phys. B441 (1995)
12.

\bibitem{HERA} I. Abt {\it et al.}, The H1 Collaboration, Nucl. Phys. B107
(1993) 515; 
M. Derrick {\it et al.}, The ZEUS Collaboration, Phys. Lett. B316 (1993)
412. 

\bibitem{NMCdeut} P. Amaudruz {\it et al.}, The NMC, Phys. Rev. Lett. 66
(1991) 2712; Phys. Lett. B295 (1992) 159. 

\bibitem{GRV} M. Gl\"uck, E. Reya, and A. Vogt, Z. Phys. C48 (1990) 471;
Z. Phys. C67 (1995) 433.

\bibitem{dglap} V.N. Gribov and L.N. Lipatov, Sov. J. Nucl. Phys. 15
(1972) 438; {\it ibid}, 675;
Yu.L. Dokshitzer, Sov. Phys. JETP 46 (1977) 641;
G. Altarelli and G. Parisi, Nucl. Phys. B 126 (1977) 298. 

\bibitem{Jaffe} R. L. Jaffe, Phys. Rev. Lett. 50 (1983) 228.

\bibitem{CRR} F. E. Close, R. G. Roberts, and G. G. Ross, Phys. Lett.
B129 (1983) 346.

\bibitem{EMC} J. Ashman {\it et al.}, EMC, Nucl. Phys. B.238 (1989) 1; 
D. Adams {\it et al.}, SMC, Phys. Lett. B329 (1994) 399; Erratum  B339
(1994) 332; 
K. Abe {\it et al.}, SLAC-E143, Phys. Rev. Lett. 74 (1995) 346, and
preprints SLAC-PUB-94-6508 and SLAC-PUB-95-6734. 

\bibitem{DI} D. Indumathi, University of Dortmund Preprint DO-95/21,
1995, to appear in Phys. Lett. B. 

\bibitem{MRS} A.D. Martin, R.G. Roberts, and W.J. Stirling, Phys. Rev.
D50 (1994) 6743.

\bibitem{CTEQ} J. Botts {\it et al.}, The CTEQ collaboration, Phys. Lett.
B304 (1993) 159. 

\bibitem{Sullivan} J. D. Sullivan, Phys. Rev. D5 (1972) 1732.

\bibitem{GRVpol} M. Gl\"uck, E. Reya, and W. Vogelsang, Phys. Lett.
B359 (1995) 201. 

\bibitem{JCRR} R.L. Jaffe, F. E. Close, R. G. Roberts, and G. G. Ross,
Phys. Lett. B134 (1984) 449, Phys. Rev. D31 (1985) 1004. 

\bibitem{Zhu} W. Zhu and J. G. Shen, Phys. Lett. B219 (1989) 107; 
W. Zhu and L. Qian, Phys. Rev. C45 (1992) 1397.

\bibitem{LS} M. Ericson and A. W. Thomas, Phys. Lett. B128
(1983) 112;
C. H. Llewellyn Smith, Phys. Lett. B128 (1983) 107;
S.A. Akulinichev, S.A. Kulagin, and G.M. Vagradov, Phys. Lett.
B158 (1985) 485.

\bibitem{FNAL} D.M. Alde, {\it et al.}, The E772 Collaboration,
Phys. Rev. Lett. 64 (1990) 2479.

\bibitem{Weiz} C. F. von Weizs\"acker, Z. Phys. 96 (1935) 431; 
H. A. Bethe and R. F. Bacher, Rev. Mod. Phys. 8 (1936) 82.

\bibitem{rad} The radii and densities, $(R_A~ \hbox{fm}, 
\rho~ \hbox{fm}^{-3})$, of the various nuclei, as quoted in \cite{EMC2}, 
are (2.6, 0.04) for ${}^{6}$Li, (2.5, 0.09) for ${}^{12}$C,
and (3.5, 0.11) for ${}^{40}$Ca. 

\bibitem{NMCsnc} NMC preliminary data on Sn/C; A. M\"ucklich, PhD.
Thesis, Ruprecht-Karls-Universit\"at, Heidelberg, 1995. 

\bibitem{spind} W. Melnitchouk, G. Piller, and A.W. Thomas, Phys. Lett.
B346 (1995) 165;
L.D. Kaptari {\it et al.}, Phys. Lett. B321 (1994) 271;
H. Khan and P. Hoodbhoy, Phys. Lett. B298 (1993) 181;
M.V. Tokarev, Phys. Lett. B318 (1993) 559;
B. Badelek and J. Kwiecinski, Nucl. Phys. B370 (1992) 278;
L.L. Frankfurt and M.I. Strikman, Nucl. Phys. B405 (1983) 557.

\bibitem{Wolo} R.M. Woloshyn, Nucl. Phys. A496 (1989) 749;
C. Ciofi degli Atti, E. Pace, G. Salme, Phys. Rev. C46 (1992) R1591. 

\bibitem{CSPS} C. Ciofi degli Atti, S. Scopetta, E. Pace, G. Salme,
Phys. Rev. C48 (1993) R968. 

\bibitem{FGS} L.L. Frankfurt, V. Guzey, and M.I. Strikman, Tel Aviv
University, Israel, preprint 1996, hep-ph9602301. 

\bibitem{Kumano} R. Kobayashi, S. Kumano, and M. Miyama, Phys. Lett.
B354 (1995) 465.

\bibitem{DIA} D. Indumathi, and M.V.N.~Murthy, 
{\it Nuclear effects on parton densities in deep inelastic
lepton hadron scattering}, Extended Abstract, DAE Symposium on Nuclear
Physics, Aligarh, India, 1989. 

\bibitem{FSL} L.L. Frankfurt, M.I. Strikman, S. Liuti, Phys. Rev. Lett.
65 (1990) 1725. 

\end{thebibliography}
\end{document}